\documentclass[]{aa}
\usepackage[varg]{txfonts}
\usepackage{lineno} 
\usepackage{amssymb,amsmath}
\usepackage{graphicx}
\usepackage{xcolor,natbib}
\usepackage{multirow}
\usepackage[T1]{fontenc}
\usepackage{ae,aecompl}
\usepackage{newtxtext, newtxmath}
\usepackage{float}
\usepackage{subfigure}
\usepackage{longtable}
\usepackage{enumitem}
\usepackage{longtable,listings}
\usepackage[flushleft]{threeparttable}
\usepackage{parcolumns}
\usepackage{hyperref}
\leftlinenumbers               
\linenumbers
\nolinenumbers
\usepackage{CJK}
\usepackage{upgreek}

\begin{document}

\title{AT2019aalc: an obscured tidal disruption event candidate in an active galactic nucleus revealed by its first flare?
}

\titlerunning{AT2019aalc in AGN}

\author{Ying Gu \inst{\ref{inst1},\ref{inst2}}
\and Xiao Li\inst{\ref{inst2}}
\and Xue-Guang Zhang\inst{\ref{inst2}}
\and En-Wei Liang\inst{\ref{inst2}}}

\institute{College of Physics and Electronic Engineering, Northwest Normal University, Lanzhou, China \label{inst1}
\and
Guangxi Key Laboratory for Relativistic Astrophysics, School of Physical Science and Technology, Guangxi University, Nanning 530004, China \label{inst2} \\
\email{xgzhang@gxu.edu.cn; lew@gxu.edu.cn}}
\date{Received XXX / Accepted XXX}

\abstract
{AT2019aalc is considered a repeating tidal disruption event (TDE) candidate occurring in an active galactic nucleus (AGN). In this paper, we highlight previously unnoticed but intriguing features that include a brief optical dimming prior to the rise of the main flare and a significantly lower post-flare luminosity compared to the pre-flare level. By applying a time-dependent obscured TDE model to AT2019aalc, we successfully reproduce its light curves with the tidal disruption of a $1.122_{-0.176}^{+0.246}\rm M_\odot$ main-sequence star by a $1.862_{-0.259}^{+0.255} \times 10^7{\rm M_\odot}$ supermassive black hole (SMBH). Both the pre-flare dimming and the reduced post-flare luminosity are attributed to obscuration of the intrinsic AGN emission. The total energy of the event derived from our fit is $1.24\times10^{51}$ ergs and about 0.2 ${\rm M_\odot}$ of debris mass is accreted by the SMBH. In addition, the $g-r$ color of the net flare in AT2019aalc exhibits a continuous reddening trend during the first flare, consistent with a gradual increase in $E(B-V)$ predicted by the obscured TDE model. These results suggest that AT2019aalc can be considered a second member of the recently proposed class of obscured TDE candidates hosted in AGNs. }
\maketitle

\section{Introduction}
When a hapless star passes too close to a supermassive black hole (SMBH), it will be torn apart by tidal forces in what is known as a tidal disruption event (TDE). Approximately half of the stellar debris becomes unbound, while the other half forms an accretion flow that remains bound to the SMBH. The circularization and accretion of the bound debris produce a bright multi-wavelength flare lasting from months to years \citep{1975Natur.254..295H,1988Natur.333..523R}. 
The characteristic light-curve behavior predicted for TDEs involves a rapid rise followed by a slower decline roughly following $t^{-5/3}$ after the peak, which eventually flattens and returns to the pre-flare level (e.g., \citealt{1988Natur.333..523R,2019GReGr..51...30S}).

The development of high-cadence, wide-field public sky surveys has enabled
the discovery of $\gtrsim$ 100 candidate events (e.g., \citealt{2021ARA&A..59...21G,2023ApJ...942L..33W,2023ApJ...955L...6Y}). Interestingly, researchers have noted significant diversity in the light-curve behavior of TDEs, which often deviates from the expectations of canonical models. For example, some TDEs show plateaus, rebrightening bumps, or reflaring phenomena in the post-peak luminosity evolution (e.g., \citealt{2019ApJ...871...15J,2024A&A...692A.262S,2023ApJ...955L...6Y}).  
These deviations may arise from several mechanisms, such as TDEs occurring in SMBH binaries (e.g., \citealt{2017MNRAS.465.3840C,2020NatCo..11.5876S}) or delayed accretion disk formation (e.g., \citealt{2022ApJ...925...67L,2025ApJ...979..235G}). 

Recent advances in time-domain astronomy have led to the discovery of numerous TDE candidates in active galactic nuclei (AGNs) (e.g. \citep{2022MNRAS.516L..66Z,2024A&A...692A.262S,2025ApJ...982..150S,2025MNRAS.537...84G}). In particular, several extreme nuclear transients in AGNs with uncertain physical origin may in fact be associated with exotic TDE scenarios. For instance, the changing-look AGN 1ES 1927+654 \citep{2019ApJ...883...94T} exhibited an X-ray drop of two orders of magnitude after its UV/optical outburst, which has been suggested to be related to a TDE \citep{2020ApJ...898L...1R,2021ApJS..255....7R,2023MNRAS.526.2331C}. Similarly, PS16dtm, an atypical TDE in a narrow-line Seyfert 1 (NLS1) galaxy, features a $\sim 100$-day light-curve plateau attributed to Eddington-limited accretion and stellar debris obscuration \citep{2017ApJ...843..106B}. Furthermore, \citet{2021ApJ...920...56F} systematically analyzed a class of rapid flares occurring in NLS1s, suggesting that some of these transients may be powered by TDEs. More recently, another peculiar TDE candidate, CSS100217, which also occurred in an NLS1, exhibits a post-flare brightness that is significantly lower than its pre-flare level. \cite{2025A&A...702L...8G} proposed a time-dependent obscured TDE model to explain this unique variability, in which a $4.7\,\rm M_\odot$ main-sequence star is tidally disrupted by a $1.17 \times 10^7\,\rm M_\odot$ supermassive black hole.

AT2019aalc is a dramatic transient discovered in the nucleus of a Seyfert 1 AGN at $z=0.0356$. The source displays two major optical flares separated by approximately four years. The first flare in early 2019 was interpreted as a TDE candidate \citep{2021TNSTR3680....1V,2024MNRAS.529.2559V}, while the subsequent significant rebrightening observed in 2023 has further been interpreted as a possible repeating TDE candidate \citep{2023TNSAN.194....1V,2026A&A...706A.324V}. An alternative explanation attributes the flares of AT2019aalc to radiation pressure instabilities in the inner part of the accretion disk \citep{2025ApJ...989..173S}. In particular, AT2019aalc exhibited peculiar dimming behavior in the pre- and post-flare phases of the first optical main flare: a brief dimming occurred during the $\sim 100$~days prior to the flare rise around MJD-53000 $\sim$ 5550 days, and the post-flare luminosity dimmed again below the pre-flare baseline around MJD-53000 $\sim$ 6100 days. For the $g$ ($r$) band, both dips reached a depth of $\sim 0.2~\mathrm{mag}$ ($\sim 0.1~\mathrm{mag}$).

In this work, motivated by the peculiar light-curve behavior of AT2019aalc, we investigate whether a time-dependent obscured TDE model can account for these observations. It should be noted that, due to the limited follow-up data for the second flare, 
only the first flare of AT2019aalc is considered in the light-curve analysis. Although the post-flare phase of the second flare exhibits a clear declining trend, its overall luminosity remains above the pre-first-flare level.
The manuscript is organized as follows. In Sect. \ref{appobs}, we present an obscured TDE model and apply this model to the light curve of AT2019aalc. The analysis of optical color variation is given in Sect. \ref{color}. The discussions and conclusions are given in Sects. \ref{dis} and \ref{con}, respectively. Throughout the 
manuscript, we have adopted the cosmological parameters of $H_{0}$=70 km s$^{-1}$ Mpc$^{-1}$, $\Omega_{m}$=0.3, and 
$\Omega_{\Lambda}$=0.7.

\begin{figure*}\label{LC}
\centering
\begin{minipage}[c]{0.49\textwidth}
  \centering  \includegraphics[width=\textwidth]{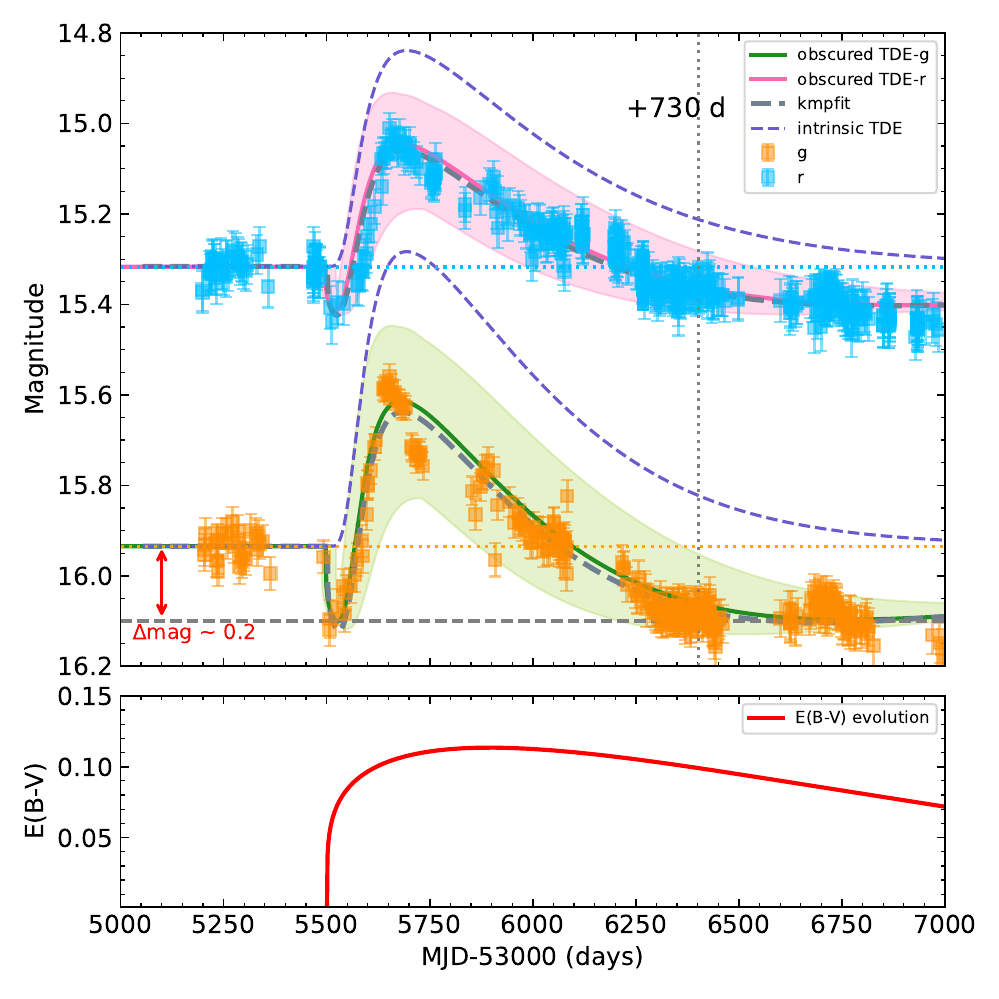}
\end{minipage}
\hfill
\begin{minipage}[c]{0.49\textwidth}
  \centering
   \includegraphics[width=\textwidth]{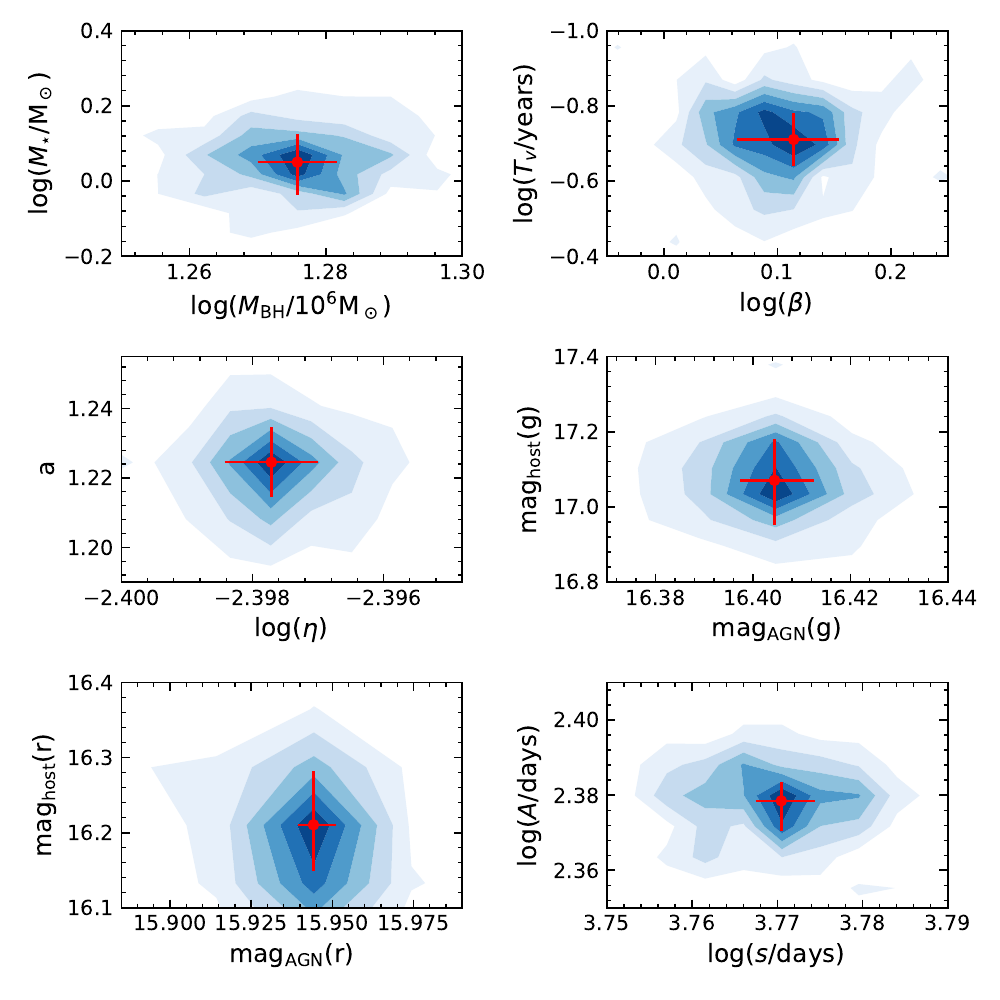}
\end{minipage}
\vspace{-0.3cm}
\caption{Left panels: the top panel shows the ZTF single-exposure photometric data in the \emph{g}- and \emph{r}-bands are shown as orange and blue points, respectively. The best-fit obscured TDE model obtained from a simultaneous fit to both bands is displayed as solid curves, with the shaded regions indicating the corresponding confidence intervals derived from the uncertainties of the model parameters.For comparison, the dashed curves represent the intrinsic TDE light curves recovered by removing the obscuration component. The horizontal dotted lines mark the mean pre-flare magnitudes in each band. The gray dashed line represents the mean post-flare magnitude, which is about 0.2 mag fainter than the pre-flare level (red arrow) in g-band. The vertical gray dotted line marks the time of the spectroscopic observation. The bottom panel show the time evolution of the color excess $E(B-V)$ inferred from the best-fit parameters of the obscured TDE model. Right panels: MCMC-determined two-dimensional distributions of obscured TDE model parameters. In each right panel, the solid circle plus error bars in red marks the accepted values and the corresponding uncertainties of the model parameters.}
\label{Fig_1}
\end{figure*}

\section{Application of the Obscured TDE Model to AT2019aalc}\label{appobs}
We retrieved the AT2019aalc historical $g$- and $r$-band light curves from the Zwicky Transient Facility (ZTF; \citealt{2019PASP..131a8002B}) forced photometry service and converted these differential fluxes into apparent magnitudes following the standard instructions provided in \cite{2023arXiv230516279M}.
The magnitudes were then corrected for foreground extinction using the maps of \citet{2011ApJ...737..103S}, assuming a Galactic extinction law with $R_V = 3.1$ and $E(B-V) = 0.039$ \citep{1989ApJ...345..245C}. The data span MJD-53000 = 5200 $\sim$ 7000 days (2018 March 23 to 2023 October 21), as shown in the top-left panel of Fig. \ref{Fig_1}. 
For clarity, all light-curve descriptions below are given with respect to MJD-53000.
It was clear that both the $g$- and $r$-band light curves exhibit lower luminosity levels during 5450 $\sim$ 5550 and 6100 $\sim$ 7000 days compared to the pre-flare phase (5200 $\sim$ 5450 days). In addition, the light curves of both bands also show a smooth declining trend during 5750 $\sim$ 7000 days that resembles the light‑curve features expected from a TDE. We therefore apply an obscured TDE model to account for the unique variability features. 

Similar to what we have recently done in \citealt{2025A&A...702L...8G}, we considered time-dependent, viscously delayed accretion-rate templates expected from the obscured TDE model, applied within a blackbody photosphere framework. Based on the fallback rates $\dot M_{fb}(t)$ from TDEFIT/MOSFIT for the tidal disruption of a $M_{*}=1{\rm M_\odot}$ main-sequence star by a $M_{\rm BH}=10^{6}{\rm M_\odot}$ black hole (BH), together with the viscous-delayed effects from stream circularization and disk accretion \citep{2013ApJ...767...25G, 2019ApJ...872..151M}, we constructed standard templates of the viscous-delayed accretion rate $\dot{M}{_a}(t_a|T{_v},\beta)$ for different impact parameters $\beta$ (the ratio of the tidal disruption radius $R_t$ to the pericenter radius $R_p$) and viscous delay times $T_{v}$. We also derived the corresponding time $t_{a}(\dot M_a|T_{v},\beta)$. For typical TDEs, the observer-frame accretion rate and time are obtained by the following scaling relations \citep{2019ApJ...872..151M}:
 \begin{equation}
        \dot{M} \propto M_{\rm BH,6}^{-1/2}\! M_{*}^{2}\! R_{*}^{-3/2}\! \dot{M}_{a}(t_a|T_{v},\beta)\;
\end{equation}
\begin{equation}
    t\propto (1+z) M_{\rm BH,6}^{1/2}\! M_{*}^{-1}\! R_{*}^{3/2}\! t_{a}(\dot M_a|T_{v},\beta)\;,
\end{equation}
where $M_{\rm BH,6}=M_{\rm BH}/10^{6}{\rm M_\odot}$,  $M_{*}$ and $R_{*}$ are the stellar mass and radius in units of ${\rm M_\odot}$ and ${\rm R_{\odot}}$, respectively, and $z$ is the redshift of the TDE host galaxy.
In addition, we adopted the main-sequence mass–radius relation from \cite{1996MNRAS.281..257T}. It should be noted that these scaling relations were derived for typical TDEs. For TDEs occurring in AGNs, the initial disruption and fallback timescales are mainly determined by the SMBH and stellar properties, while the subsequent debris evolution can be affected by AGN disk interactions, depending on the disk density (e.g., \citealt{2019ApJ...881..113C,2024MNRAS.527.8103R}). In the low-density regime, AGN-TDEs can still resemble naked TDEs \citep{2024MNRAS.527.8103R}. In addition, interactions between the AGN accretion disk and the debris stream may affect the circularization of the bound debris. Hydrodynamical simulations have shown that stream self-crossing and shock dissipation are important processes in debris circularization, with the circularization timescale and associated energy dissipation depending on the encounter parameters and radiative cooling \citep{2016MNRAS.455.2253B,2016MNRAS.461.3760H}. These processes may therefore further affect the subsequent accretion and radiative evolution of the TDE. Following \citet{2019ApJ...872..151M}, we do not explicitly model these complex circularization processes, but instead account for the overall time delay associated with circularization and subsequent accretion through an effective viscous-delay timescale $T_v$. Therefore, we apply these relations to AT2019aalc for simplicity.

Following \cite{2025A&A...702L...8G},
the observed flux of the obscured TDE in the rest frame is
\begin{equation}
F^{\rm obs}_{\lambda}(t) = \frac{2\pi hc^2}{\lambda^5} \frac{1}{e^{hc/(k\lambda T_p(t))}-1} \left[\frac{R_p(t)}{D(z)}\right]^2 \times  b_{\lambda}(t)\;,
\end{equation}
with a radius of the photosphere of
\begin{equation}
R_p(t) =  R_0 \times \left[ G M_{\rm BH} \left( {t_p}/{2\pi} \right)^2 \right]^{1/3} \left({L}/{L_{\rm Edd}} \right)^{l_p}\;,
\end{equation}
an effective temperature of
\begin{equation}
T_p(t) = \left[ {L/}{(4\pi \sigma_{SB} R_p^2}) \right]^{1/4} \;
\end{equation}
and a wavelength-dependent extinction of
\begin{equation}
b_{\lambda}(t)=10^{-0.4k(\lambda)E(B-V)(t)}\;.
\end{equation}
Here, $c$ is the speed of light, $k$ as the Boltzmann constant, and $D(z)$ is the luminosity distance at redshift $z$. $R_p(t)$ ranges 
from the minimum innermost stable circular orbit radius to maximum semimajor axis of the accreting mass. $L = {\eta\dot{M}(t)c^2}$ is the 
bolometric luminosity, $L_{\rm Edd} = 1.3\times10^{38}(M_{\rm BH}/M_\odot)\rm \;erg\;s^{-1}$ is the Eddington luminosity, $\eta<0.4$ 
is the energy transfer efficiency \citep{2014ApJ...783...23G, 2019ApJ...872..151M},  $l_p$ is the power-law exponent,  $t_p$ is the time of the peak accretion rate, $\sigma_{\rm SB}$ 
is the Stefan-Boltzmann constant, $E(B-V)(t)$ is the color excess, and $k(\lambda)$ is the extinction curve.  
We adopted the Galactic F99 extinction law  \citep{1999PASP..111...63F,Gordon2024} to model the obscuration and assumed the time‑dependent $E(B-V)$ follows a Weibull distribution of the form \citep{2025A&A...702L...8G}
\begin{equation}
E(B-V)(t) = A \cdot \left( \frac{a}{s} \right) \left( \frac{t - t_0}{s} \right)^{a-1} \exp\left[ - \left( \frac{t - t_0}{s} \right)^a \right],
\end{equation}
where \( A \) is the normalization factor, \( a \) is the shape parameter, \( s \) is the scale parameter, and \( t_0 \) is the onset time of extinction.

Finally, the apparent magnitudes observed of the obscured TDE are obtained by convolving the observer-frame spectrum, $F^{\rm obs}_\lambda(t)$, with the transmission curve of the corresponding photometric filter. Additionally, the observed apparent magnitude of the AGN is corrected using the relation
\begin{equation}
    \rm mag^{obs}_{AGN}(\lambda) = {\rm { mag_{AGN}}(\lambda)} + A_\lambda \,, 
\end{equation}
where $A_\lambda$ represents the extinction in the $\lambda$ band, and $\rm mag_{AGN}(\lambda)$ is the intrinsic apparent magnitude of the AGN in the $\lambda$ band. 
We derived $A_\lambda$ by interpolating the F99 extinction curve \citep{1999PASP..111...63F}, which provides $A_\lambda/A_V$, with $A_V = R_V E(B-V)(t)$ and $R_V = 3.1$.

We applied the obscured TDE model to the $g$- and $r$-band light curves of AT2019aalc, assuming that obscuration begins immediately at the onset of the flare. 
The model parameters were derived using the maximum likelihood method combined with the Markov chain Monte Carlo (MCMC) technique \citep{2013PASP..125..306F}, leading to the determined model parameters with a polytropic index of $\gamma=4/3$ being:
$\log(M_{\rm BH}/10^6 \rm M_\odot)=1.270_{-0.006}^{+0.006}$, 
$\log(M_{*}/\rm M_\odot)=0.050_{-0.074}^{+0.086}$, 
$\log(\beta)=0.114_{-0.044}^{+0.054}$, 
$\log(T_{v}/\text{years})=-0.718_{-0.070}^{+0.070}$, 
$\log(\eta)=-2.397_{-0.001}^{+0.001}$,  
$\log(R_{0})=-0.672_{-0.010}^{+0.009}$,  
$\log(l_{p})=-0.746_{-0.005}^{+0.001}$,  
$\log (A/\text{days})=2.380_{-0.005}^{+0.008}$, 
$a=1.229_{-0.008}^{+0.014}$, 
$\log(s/\text{days})=3.770_{-0.004}^{+0.008} $, 
$\rm mag_{AGN}(g)=16.404_{-0.008}^{+0.071}$,
$\rm mag_{host}(g)=17.059_{-0.129}^{+0.119}$ 
and $\rm mag_{AGN}(r)=15.944_{-0.007}^{+0.005}$, 
$\rm mag_{host} (r)=16.212_{-0.072}^{+0.061}$.  
The best-fitting results and the 1$\sigma$ confidence bands derived from the uncertainties of the model parameters are shown in the top-left panel of Fig. \ref{Fig_1}.
The light curves of AT2019aalc can be explained by an obscured TDE resulting from the tidal disruption of a $1.122_{-0.176}^{+0.246}\rm M_\odot$ main-sequence star by a SMBH of $1.862_{-0.259}^{+0.255} \times 10^7{\rm M_\odot}$.
The total energy of the event derived from our fit is $1.24\times10^{51}$ ergs. About 0.2 ${\rm M_\odot}$ of the debris is accreted by the central SMBH. 

The intrinsic TDE flare without obscuration is represented by the dashed purple line in the top-left panel of Fig. \ref{Fig_1}, while the color excess $E(B-V)$ is displayed in the bottom-left panel. 
In the presence of sustained obscuration,
the observed light-curve evolution can be divided into three stages corresponding to different dominant components. 
The intrinsic TDE flare starts to rise at $\sim 5500$ days, but the observed luminosity shows a brief $\sim 100$-day dimming due to obscuration of the AGN-dominated component. Subsequently, the TDE component rises to a peak around $\sim 5750$ days and slowly decays during the TDE-dominated phase, with obscuration reducing its observed luminosity. After $\sim 6200$ days, the observed emission becomes AGN-dominated again and obscuration keeps the observed luminosity below the pre-flare level.

\begin{figure}
\includegraphics[width=1\columnwidth]{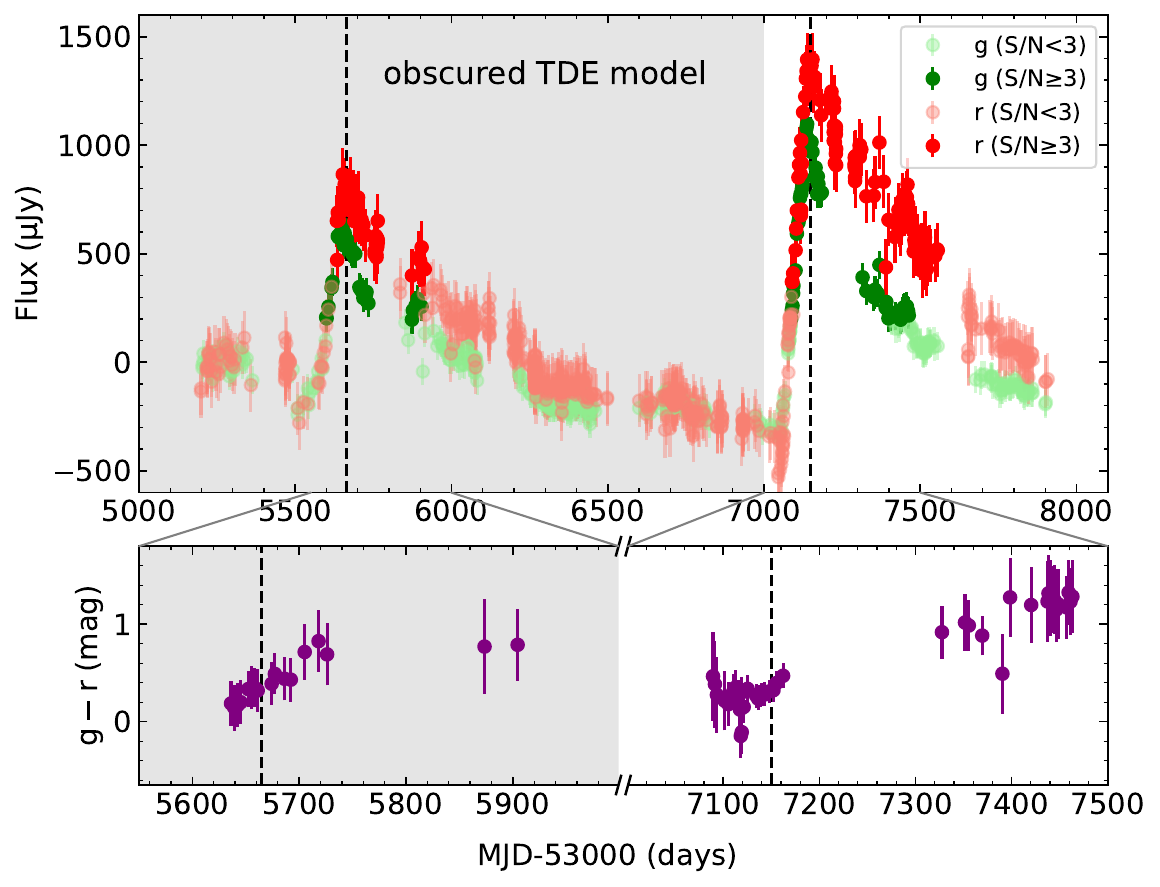}
 \centering
\caption{Top panel: 
net flare flux in the ZTF $g$-band (green) and $r$-band (red), obtained after subtracting the average flux in each band during the pre-flare phase (5200 $\sim$ 5400 days, see Fig. \ref{Fig_1}).
Bottom panel: 
$g-r$ color evolution of AT2019aalc for data with $F_{\rm flare}/\sigma_{\rm flare}$ (S/N) greater than 3. The gray shaded region marks the first flare used for fitting with the obscured TDE model and the corresponding color evolution.
}\label{Fig_2}
\end{figure}

\section{Analysis of optical color variation of AT2019aalc}\label{color}
The color evolution is crucial for understanding the physical properties of TDE flares. In general, optical TDEs are characterized by a persistent blue continuum with only weak color evolution \citep{2017ApJ...842...29H,2021ApJ...908....4V}, which is commonly used to distinguish TDEs from variable AGNs and supernovae \citep{2021ARA&A..59...21G}. 
The red optical color evolution of AT2019aalc during its two major optical flares was reported by \cite{2025ApJ...989..173S}. However, the contributions of the host galaxy and the AGN to the observed light-curve variability during the flares were neglected in their analysis.
Therefore, we reanalyzed the color evolution of AT2019aalc.   
We collected ZTF data for AT2019aalc spanning 5300 $\sim $ 7408 (2018 March 21 to 2025 August 21). We adopted the average fluxes in the $g$ and $r$ bands during the pre‑flare phase (5200 $\sim$ 5400 days, see Fig. \ref{Fig_1}) as the baseline fluxes ($F_{\rm base}$), and used the corresponding standard deviations as the baseline uncertainties ($\sigma_{\rm base}$). The net flare flux in each band ($F_{\rm flare}$) was then obtained by subtracting the baseline flux from the observed flux. The uncertainty of the net flare flux was calculated as ${\sigma _{\rm flare}} = \sqrt {\sigma _{\rm obs}^2 +  \sigma _{\rm base}^2}$. Finally, the $g-r$ color was constructed using only data  with $F_{\rm flare}/\sigma_{\rm flare}$ ($\rm S/N$) greater than 3 in both the $g$ and $r$ bands.

The $g$- and $r$-band light curves of the net flare are shown in the upper panel of Fig. \ref{Fig_2}, while the corresponding $g-r$ color evolution is presented in the lower panel.  
The $g-r$ color exhibits a continuous reddening trend during 5620 $\sim$ 5760 days.
This behavior is consistent with the obscured TDE model, which predicts that $E(B-V)$ increases from 5500 to 5900 days. Our model further predicts that $E(B-V)$ will gradually decrease after 5900 days, leading to a reduction in the obscuration-induced $g-r$ color. However, the analysis of the color evolution for the second flare shows a more pronounced reddening trend. If this reddening trend is caused by obscuration, additional obscuring material would be required.

\section{Discussions}\label{dis}
AT2019aalc has been proposed as a repeating TDE candidate \citep{2023TNSAN.194....1V,2026A&A...706A.324V}. Owing to the limited data for the second flare, we restricted our analysis to the first flare. The inferred impact parameter is $\beta \sim 1.30$.
Hydrodynamical simulations suggest that, for a polytropic index of $\gamma=4/3$, such an encounter may result in a partial tidal disruption, leaving a surviving stellar core that can undergo further partial disruptions and produce subsequent flares \citep{2013ApJ...767...25G,2017A&A...600A.124M,2020ApJ...905..141L,2024MNRAS.532...89S}. However, it should be noted that partial TDEs are expected to exhibit a steeper light-curve decline ($\sim t^{-9/4}$) than classical TDEs ($\sim t^{-5/3}$) \citep{2019ApJ...883L..17C}. The partial TDE candidate AT2022dbl even exhibits a slightly steeper decline, following a power law of approximately $t^{-2.6}$ \citep{2025ApJ...987L..20M}.  In contrast, as shown in Fig.~\ref{Fig_1}, the first flare of AT2019aalc is consistent with the canonical TDE decline behavior, although the obscuration effect included in our model would generally further steepen the observed decline. Furthermore, the fallback timescale estimated from our model  
is $\sim160$ days, while the inferred viscous timescale $T_v$ is $\sim66$ days, suggesting that viscous delays are unlikely to dominate the light-curve decline. Therefore, the observed decline does not provide strong evidence in favor of a partial disruption scenario.

\begin{figure}
\includegraphics[width=1\columnwidth]{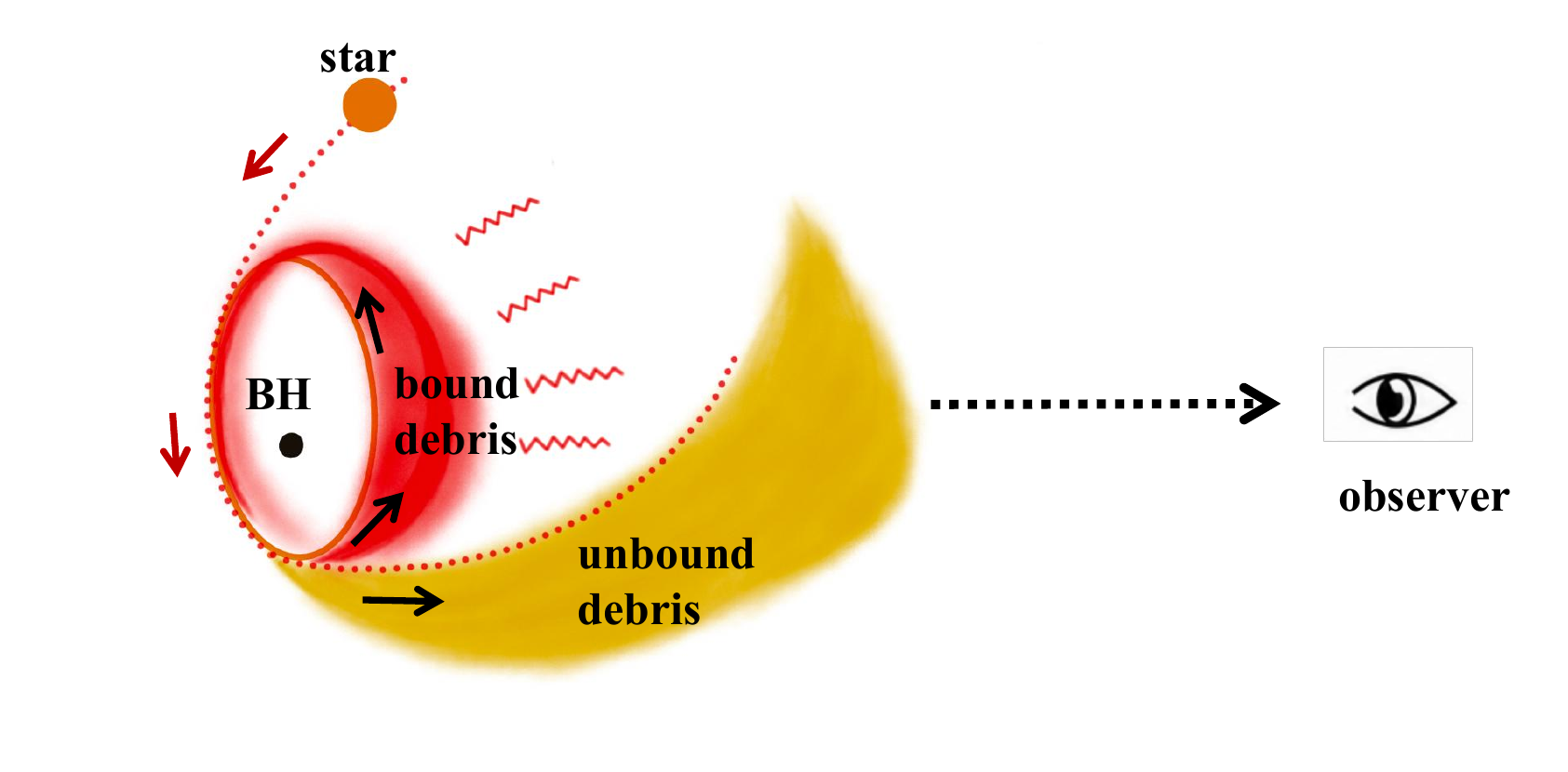}
 \centering
\caption{Schematic illustration of obscuration by unbound stellar debris in a TDE. The bound debris circularizes into an accretion disk and is subsequently accreted onto the SMBH, producing a TDE flare. Inhomogeneous unbound debris causes line-of-sight obscuration, leading to wavelength-dependent extinction of both AGN and TDE emission. The components are schematic and not drawn to scale.}\label{cartoon}
\end{figure}

\begin{figure}
\includegraphics[width=1\columnwidth]{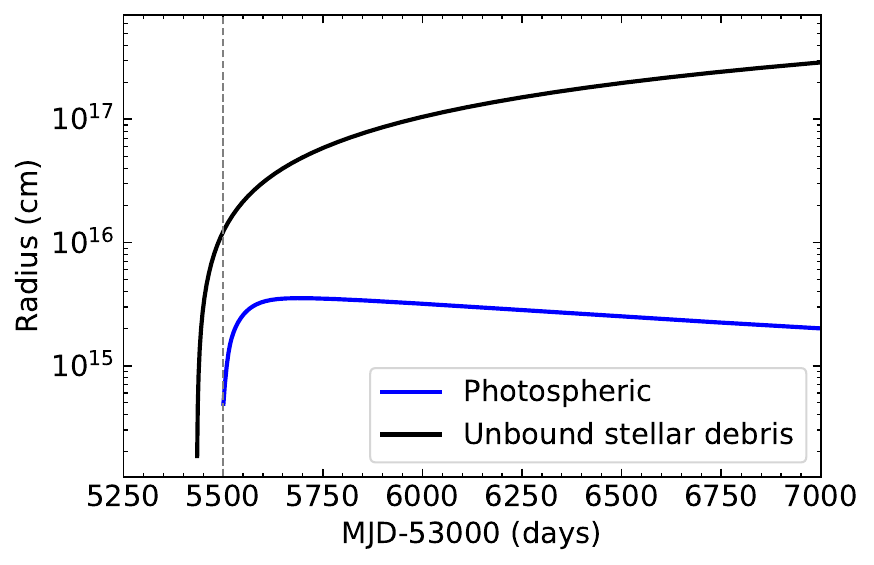}
 \centering
\caption{Time evolution of the radius of the unbound stellar debris and the photospheric radius of the TDE. The dashed gray line marks the start time of the flare. }\label{Fig_3}
\end{figure}

An alternative scenario is that the observed double-flare behavior may arise from the sequential tidal disruption of a binary stellar system by the SMBH, with the two stars disrupted at different times  \citep{2015ApJ...805L...4M,10.1093/mnrasl/slaf080,2026ApJ...997..186Z}. Each individual disruption of a binary member is expected  to resemble a canonical TDE. In this scenario, the unbound stellar debris from the disruption naturally provides a transient, compact, and highly anisotropic source of obscuration for the emission from the TDE flares and the intrinsic AGN activity \citep{2025A&A...702L...8G}. A schematic illustration of this scenario is shown in Fig. \ref{cartoon}. We assumed that obscuration occurs at the onset of the flare; however, because of viscous effects \citep{2019ApJ...872..151M}, the star has in fact already been tidally disrupted before the flare is produced.  
\citet{2009MNRAS.400.2070S} showed that the most energetic unbound stellar debris is ejected at speeds of $v_{\rm max}=(3R_*/R_{\rm p})^{1/2}v_{ p}$, where $v_{p}=(2GM_{BH}/{R_{p}})^{1/2}$ is the orbital velocity at pericenter.  
Therefore, at the onset of obscuration, the unbound stellar debris lies at a distance of $v_{\rm max}T_v\sim0.004$ pc from the central BH, which exceeds the TDE photospheric radius at that epoch (see Fig. \ref{Fig_3}). As time evolves, the debris radius continues to exceed the photospheric radius. The mass of the unbound debris ejected in the first flare is approximately $0.92\,\rm M_\odot$.
The unbound stellar debris during a TDE would generally be concentrated in a cone or “fan” close to the orbital plane (e.g., \citealt{1988Natur.333..523R,2009MNRAS.400.2070S,2019GReGr..51...30S}). Therefore, at maximum obscuration (about 400 days after disruption), the unbound stellar debris lies at $R \sim 0.02$ pc from the central SMBH. The number density of particles within the cone of unbound debris can be estimated as $n\sim (M/m_{p})/(R^2\Delta R\Delta\Omega/3)\sim10^{10} \;\rm cm^{-3}$, where $\Delta\Omega\sim48^{1/2}(R_*/R_{\rm p})^{3/2}$ is the solid angle subtended by the unbound stellar debris, and $\Delta R\sim R (3R_*/R_{p})^{1/2}$ is the radial dispersion of the debris at a fixed azimuthal angle \citep{2009MNRAS.400.2070S}. The density of the radial column can be estimated as $N\sim n\cdot\Delta R\sim10^{26}\;\rm cm^{-2}$. Therefore, if the unbound stellar debris lies along our line of sight (LOS), it can significantly dim the intrinsic AGN activity and the TDE flare. The stronger reddening observed during the second flare can be explained by unbound debris produced during the event responsible for it, which further enhances the LOS obscuration. Although the AGN disk environment may modify the velocity and morphology of the unbound debris, a lower debris velocity would keep the material closer to the SMBH and could even enhance the column density, leaving our obscuration argument unchanged. Another variable-obscuration scenario typically invokes moving clouds around the SMBH, which only obscure a relatively small central region of the AGN \citep{2010A&A...517A..47M,2024A&A...684A.101M}. \citealt{2010A&A...517A..47M} showed that the obscuring clouds are not spherically symmetric but rather exhibit a cometary morphology, consisting of a dense head followed by a gradually dispersing tail, which is consistent with the expectations of our model. In addition, these obscuring clouds may originate from clouds in the broad-line region (BLR) \citep{2010A&A...517A..47M}.
Assuming that the obscuring clouds move at the Keplerian velocity around the SMBH, the characteristic dynamical timescale is given by $\tau_{\rm dyn} \sim (R_{\rm cloud}^3/GM_{\rm BH})^{1/2}$,  where $R_{\rm cloud}$ denotes the distance of the obscuring cloud from the SMBH. We approximated the cloud distance as $R_{\rm cloud}=R_{\rm BLR}$, where $R_{\rm BLR}$ is the radius of BLR.
The radius of the BLR is estimated from the radius-luminosity relationship \citep{2013ApJ...767..149B} as
$R_{\rm BLR} \sim 5 \times 10^{16}\left(L_{5100}/10^{44}\ \mathrm{erg\ s^{-1}}\right)^{0.53}\ \mathrm{cm}$, where $L_{5100} = 10^{43.5}\ \mathrm{erg\ s^{-1}}$ is adopted \citep{2025ApJ...989..173S}.
This yields a characteristic timescale of $t_{\rm dyn} \sim 7$ yr, which is consistent with the timescale required by our model. In addition, our fits yield a maximum extinction of $A_V \sim 0.4$. Assuming a Milky Way-like dust-to-gas ratio \citep{2009MNRAS.400.2050G}, where
$N_{\rm H} = 2.21  \times 10^{21}\ A_V\ \mathrm{cm^{-2}}$, we obtain a hydrogen column density of the obscuring cloud along the LOS of $N_{\rm H} \sim10^{21}\ \mathrm{cm^{-2}}$.  

The SDSS spectrum of AT2019aalc was obtained in April 2008 \citep{2009ApJS..182..543A},  
while the Keck/LRIS \citep{1995PASP..107..375O} spectrum was taken approximately two years after the first flare.
However, unlike the TDE-related emission features commonly observed in TDEs, the SDSS and LRIS spectra of AT2019aalc show no significant changes in the Balmer line profiles and/or strengths \citep{2025ApJ...989..173S,2026A&A...706A.324V}. The diverse features observed in the optical spectra of TDEs have led to the establishment of several spectroscopic classes (TDE-H, TDE-He, TDE-H+He, and TDE-featureless; e.g., \citealt{2014ApJ...793...38A,2026arXiv260609325T}).
These classifications are mainly based on high signal-to-noise spectra obtained within several tens of days around the optical peak. However, TDE spectra evolve with time (e.g.,  \citealt{2019ApJ...887..218L,2020MNRAS.499..482N}), and individual events may exhibit different spectroscopic properties at different evolutionary stages \citep{2019MNRAS.488.1878N}. As the transient emission fades, TDE-related emission lines can weaken and may eventually disappear at late times (e.g., \citealt{2016MNRAS.463.3813H,2016MNRAS.455.2918H,2019ApJ...887..218L}). The LRIS spectrum of AT2019aalc was obtained $\sim$730 days after the first optical flare (the vertical gray dotted line in the upper-right panel of Fig.\ref{LC}), much later than the epochs typically used for TDE spectroscopic classification. Therefore, the observed spectrum at this stage is likely dominated by persistent AGN and host-galaxy emission, naturally explaining its similarity to the pre-flare SDSS spectrum. In addition, the unbound debris produced during the disruption may also affect the emergent spectrum through photoionization and radiative transfer processes. Previous studies have shown that unbound stellar debris can act as a reprocessing medium, with its ionized layer producing broad emission lines and contributing to the optical continuum \citep{2009MNRAS.400.2070S}. However, these transient spectral signatures depend on the strength of the central ionizing radiation and may become difficult to detect at late times as the ionizing luminosity declines. Furthermore, interactions between the AGN accretion disk and TDE debris may generate shocks and modify the energy dissipation and radiation processes, producing spectra that differ from classical TDE or AGN expectations \citep{2019ApJ...881..113C}. Since such interactions occur on timescales comparable to the debris fallback time, their associated transient spectral signatures may have already weakened or disappeared by the LRIS observation epoch.

AT2019aalc exhibits a recurring "bump" structure during its decaying phases. Similar phenomena are observed in many transient Bowen Fluorescence Flares, and TDEs (e.g., \citealt{2015MNRAS.452...69M, 2024A&A...692A.262S, 2023ApJ...955L...6Y, 2025ApJ...979..235G,2025ApJ...986..174G}). \citet{2026A&A...706A.324V} suggested that the bumps in AT2019aalc may be associated with the long-term evolution of TDE debris in an AGN environment, whereas \citet{2025ApJ...989..173S} proposed they could be driven by radiation pressure instabilities in the inner part of the accretion disk. Since the detected "bump" in our data falls within the confidence interval, further discussion is beyond the scope of this paper.

The color evolution of the two main flares is inconsistent with that of typical optical AGNs, which usually exhibit a stronger bluer-when-brighter trend \citep{2022MNRAS.510.1791N,2025MNRAS.543..121S}. A persistently red optical color ($g-r>0$) has been reported in the TDE candidate AT2022fpx and has been attributed to strong nuclear dust obscuration \citep{2025ApJ...990...22L}. 
In contrast, if the increase in $g-r$ of AT2019aalc is attributed to obscuration, it implies progressively stronger LOS obscuration, consistent with our obscured TDE model.

In our model, the effects of the AGN accretion disk environment on the TDE progenitor are not explicitly included.
We noted that stars embedded in AGN disks may either form in situ through disk gravitational instability or be captured from the surrounding nuclear cluster, with in situ formation potentially leading to a top-heavy mass function under extreme nuclear conditions \citep{2012ApJ...749..168M,2020MNRAS.493.3732D}.
Furthermore, stars embedded in AGN disks may undergo evolutionary pathways different from those in typical environments (e.g., \citealt{2021ApJ...910...94C,2021ApJ...916...48D}). Owing to the abundant gas supply in AGN disks, some stars may significantly increase their masses through accretion. However, the final stellar mass is regulated by local disk properties, including gas density, accretion efficiency, stellar location, and feedback processes. 
In some cases, stars embedded in AGN disks may instead experience prolonged main-sequence evolution due to sustained hydrogen supply and efficient internal mixing, following the “immortal star” evolutionary pathway \citep{2026ApJ...997..206X,2026arXiv260422913D}.
Moreover, the TDE progenitor does not necessarily represent the population of massive stars that have undergone long-term evolution and substantial mass growth within the AGN disk. A TDE may also originate from stars captured from the surrounding nuclear star cluster or from stars that have not experienced significant accretion growth before disruption. Therefore, the stellar mass inferred from our model should be interpreted as the mass of the star disrupted in this specific TDE event, rather than as a representative mass of the entire nuclear stellar population.

\section{Conclusions}\label{con}
We have demonstrated that the first flare of AT2019aalc can be well described by an obscured TDE, in which a main-sequence star with a mass of $1.122_{-0.176}^{+0.246}\rm M_\odot$ is tidally disrupted by the SMBH with a mass of $1.862_{-0.259}^{+0.255} \times 10^7{\rm M_\odot}$. The pre-flare dimming and the reduced post-flare luminosity relative to the pre-flare level can be explained by obscuration of the intrinsic AGN emission. The gradual reddening in $g-r$ further supports a scenario of increasing extinction during the flare evolution, consistent with the expectations of the obscured TDE model. These results suggest that AT2019aalc can be considered a second member of the recently proposed class of obscured TDE candidates hosted in AGNs.

\begin{acknowledgements}
We gratefully acknowledge the anonymous referee for giving us
constructive comments and suggestions to greatly improve our
paper. We thank the participants of the TDE FORUM (Full-process Orbital to Radiative Unified Modeling) online seminar series for their inspiring discussions. We thank Yujun Yao, Xinzhe Wang for useful discussions. This work is supported by the National Natural Science Foundation of China (grants NSFC-12173020, 12373014 and 12133003). Gu gratefully thank the kind financial support from the Innovation Project of Guangxi Graduate Education (YCBZ2024007). 

\end{acknowledgements}
\bibliographystyle{aa} 
\bibliography{references} 
\end{document}